\documentclass[a4paper,11pt]{article}
\usepackage{pos}

\title{Machine learning aided noise filtration and signal classification for CREDO experiment}
 \ShortTitle{ML aided signal classification for CREDO experiment}

\author*[a]{Łukasz Bibrzycki}
\emailAdd{lukasz.bibrzycki@up.krakow.pl}
\author[a]{Olaf Bar}
\author[a]{Marcin Piekarczyk}
\author[b]{Michał Niedźwiecki}
\author[c]{Krzysztof Rzecki}
\author[d]{Sławomir Stuglik}
\author[d]{Piotr Homola}

\affiliation[a]{Pedagogical University of Krakow, Institute of Computer Science,\\
ul. Podchorążych, 30-084 Kraków, Poland}

\affiliation[b]{Cracow University of Technology, Faculty of Computer Science and Telecommunications, 31-155 Krak\'ow, Poland}

\affiliation[c]{AGH University of Science and Technology, 30-059 Krak\'ow, Poland}

\affiliation[d]{Institute of Nuclear Physics Polish Academy of Sciences, 31-342 Krak\'ow, Poland}

\forColl{CREDO} 

\abstract{The wealth of smartphone data collected by the Cosmic Ray Extremely Distributed Observatory (CREDO) greatly surpasses the capabilities of manual analysis. So, efficient means of rejecting the non-cosmic-ray noise and identification of signals attributable to extensive air showers are necessary. To address these problems we discuss a Convolutional Neural Network-based method of artefact rejection and complementary method of particle identification based on common statistical classifiers as well as their ensemble extensions. These approaches are based on supervised learning, so we need to provide a representative subset of the CREDO dataset for training and validation. According to this approach over 2300 images were chosen and manually labeled by 5 judges. The images were split into spot, track, worm (collectively named signals) and artefact classes. Then the preprocessing consisting of luminance summation of RGB channels (grayscaling) and background removal by adaptive thresholding was performed. For purposes of artefact rejection the binary CNN-based classifier was proposed which is able to distinguish between artefacts and signals. The classifier was fed with input data in the form of Daubechies wavelet transformed images.
In the case of cosmic ray signal classification, the well-known feature-based classifiers were considered. As feature descriptors, we used Zernike moments with additional feature related to total image luminance.
For the problem of artefact rejection, we obtained an accuracy of 99\%. For the 4-class signal classification, the best performing classifiers achieved a recognition rate of 88\%.}

\FullConference{37$^{\rm{th}}$ International Cosmic Ray Conference (ICRC 2021)\\
		July 12th -- 23rd, 2021\\
		Online -- Berlin, Germany}

\begin{document}
\maketitle

\section{Introduction}
The Cosmic Ray Extremely Distributed Observatory (CREDO) is an international project aimed at registering the various kinds of radiation, of both terrestrial and extraterrestrial origin~\cite{Homola_2020}. Even though CREDO is sensitive to all kinds of radiation, its particular interest is focused on the secondary radiation produced by collisions of the Ultra High Energy Cosmic Rays with the atmosphere and called Extensive Air Showers.  The CREDO computing and database infrastructure is a hub to aggregate the data collected by various detectors like Cosmic Watches~\cite{CosmicWatch} and other low cost scintillator detectors but the flagship of the project is the CREDO Detector application installed on thousands of smartphones worldwide and able to register the cosmic rays while they pass through CMOS sensors of smartphone cameras \cite{Bibrzycki_2020}. There exist a few other initiatives of this type like DECO \cite{vandenbroucke2015detecting} and CRAYFIS~\cite{whiteson2016searching}. The common feature of all these projects is conducting the observations on the planetary scale. This way they are complementary to specialized, highly efficient but still limited in space experiments like Auger~\cite{2015172}, Telescope Array~\cite{Abu_Zayyad_2013} or IceCube~\cite{AHLERS201873}.
Currently there are about $10^7$ candidate detections in the CREDO database and to obtain the scientifically significant statistics this number is planned to increase by two orders of magnitude. CREDO is a citizen science type project and its performance relies on the commitment of    the volunteer participants. 
To sustain the users' commitment CREDO utilizes various gamification strategies, eg. periodic Particle Hunters Competitions. These competitions temporarily increase the number candidate detections sent to the CREDO database but this increased activity also results in the surge of artefacts. The artefacts are images that cannot be attributed to particle's passage through the sensor but rather result from hardware malfunction or user's cheating. The artefacts can be eliminated either on-line ie. using the algorithm working on the smartphone~\cite{borisyak2017muon} or off-line ie. through the analysis and tagging the images stored in the database. The software used to filter the artefacts and pass signals is usually called the trigger. Here we discuss the off-line Convolutional Neural Network (CNN) based trigger, even though the baseline trigger we start our discussion from is both reasonably efficient and does not require extensive computational resources, so that can be used also on line with limited resources available in smartphones. Another problem we want to discuss here is the classification of signals from CMOS sensors. There is some recognized terminology in the literature in this respect, namely spots - for point-like images, tracks - for straight lines and worms - for images containing curvy-linear signals. Moreover, some of these traces can be even assigned to particular particle species, eg. thin tracks are believed to be produced by muons while worms - by $\delta$-electrons~\cite{Groom2002}. Here we refrain from discussing the relation between trace shape and particle assignment and base our class assignment exclusively upon morphological properties of traces. This is because evidence collected so far from the CREDO database suggests that details of tracks are strongly hardware and software dependent. Eg. the thickness of the track or worm depends the underlying CMOS array resolution as well as on the algorithm used to transform the raw signal into an image saved in one of the commonly used formats. We believe that reliable trace to particle species assignment requires both deep understanding of the image processing algorithm and additional data like time correlations with images registered by nearby devices. Such correlations provide the evidence of possible EAS signal dominated by muons. To perform the signal classification we test a rich array of classical statistical classifiers including kNN, SVM, tree based classifiers as well as dense feed forward networks. We also test their ensemble versions.  All these classifiers require a carefully chosen features to perform the training. The reason for which we use feature based classifiers instead of CNNs is that most of the analyses performed so far used various geometrical properties of signal to describe and classify them \cite{waniak2006removing}. Here we use more general image features expressed in terms of Zernike moments which proved their usefulness in the recognition of facial expression~\cite{6460899,LAJEVARDI20101771} or gender classification based on facial images~\cite{10.1007/978-3-319-60618-7_39}.

To train both the CNN based trigger and the statistical classifiers used for the four class hit identification we use the set of 2354 examples drawn from the CREDO database and labelled by 5 annotators.
\section{Baseline trigger}
\begin{figure}
    \centering
    \vspace{-1.cm}
    \includegraphics[scale=.9]{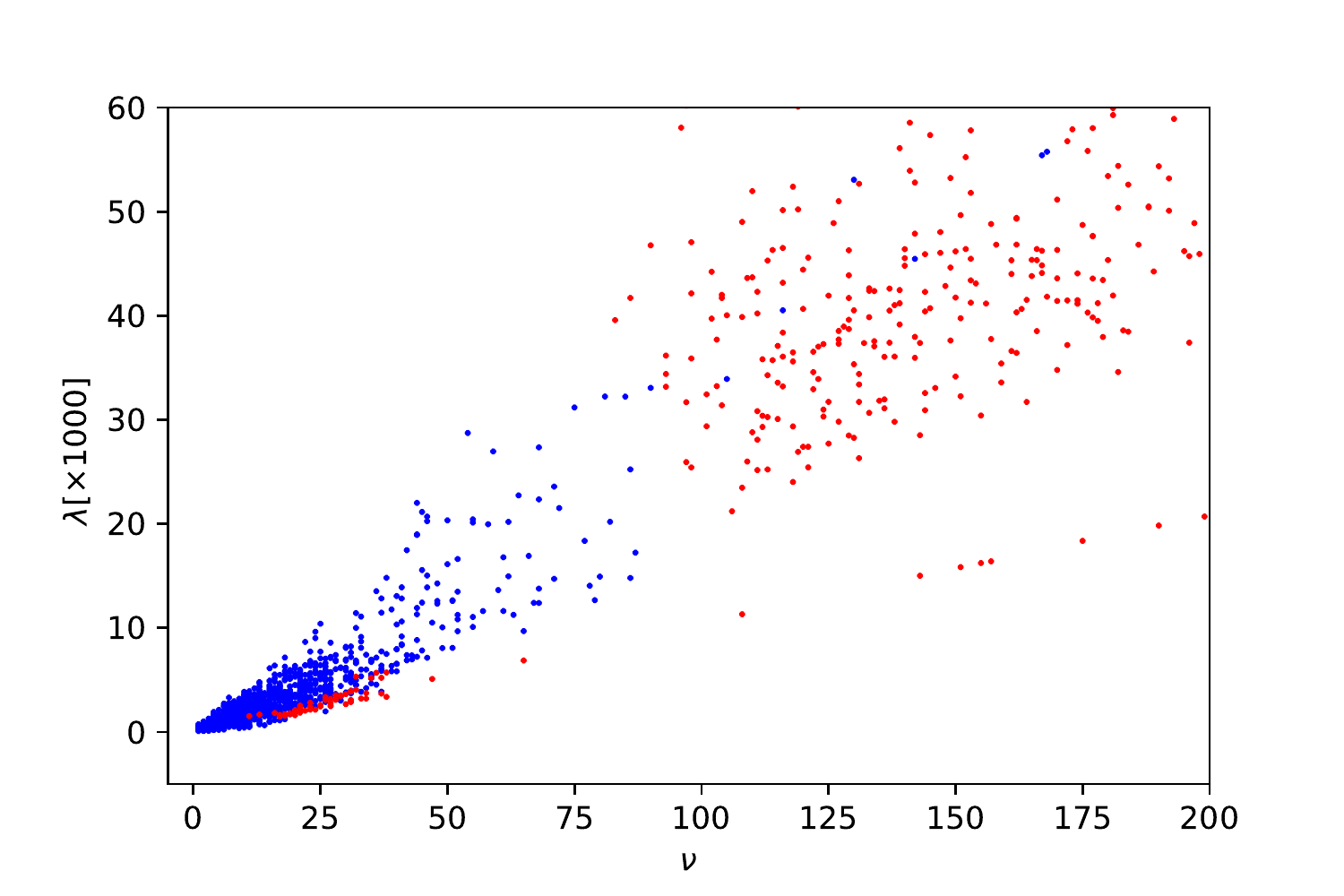}
    \vspace{-.4cm}
    \caption{Distribution of the training dataset points projected on the ($\nu,\lambda$) feature plane.}
    \label{fig:basetrigger}
\end{figure}

To begin with, we consider the baseline trigger, ie. a simple classifier which assigns the hit image to either signal or artefact class based on the most general and distinctive properties if the image. The role of the baseline trigger is twofold. First, it provides a lower bound of the classification accuracy for the more elaborate, eg. neural network based classifiers, while being entirely explainable~\cite{e23010018}. Second, relying on the simple heuristics they are much less computationally challenging, so can be effortlessly implemented as online triggers working with limited resources of smartphones. 
\begin{figure}[h]
    \centering
    \includegraphics[scale=.5,clip]{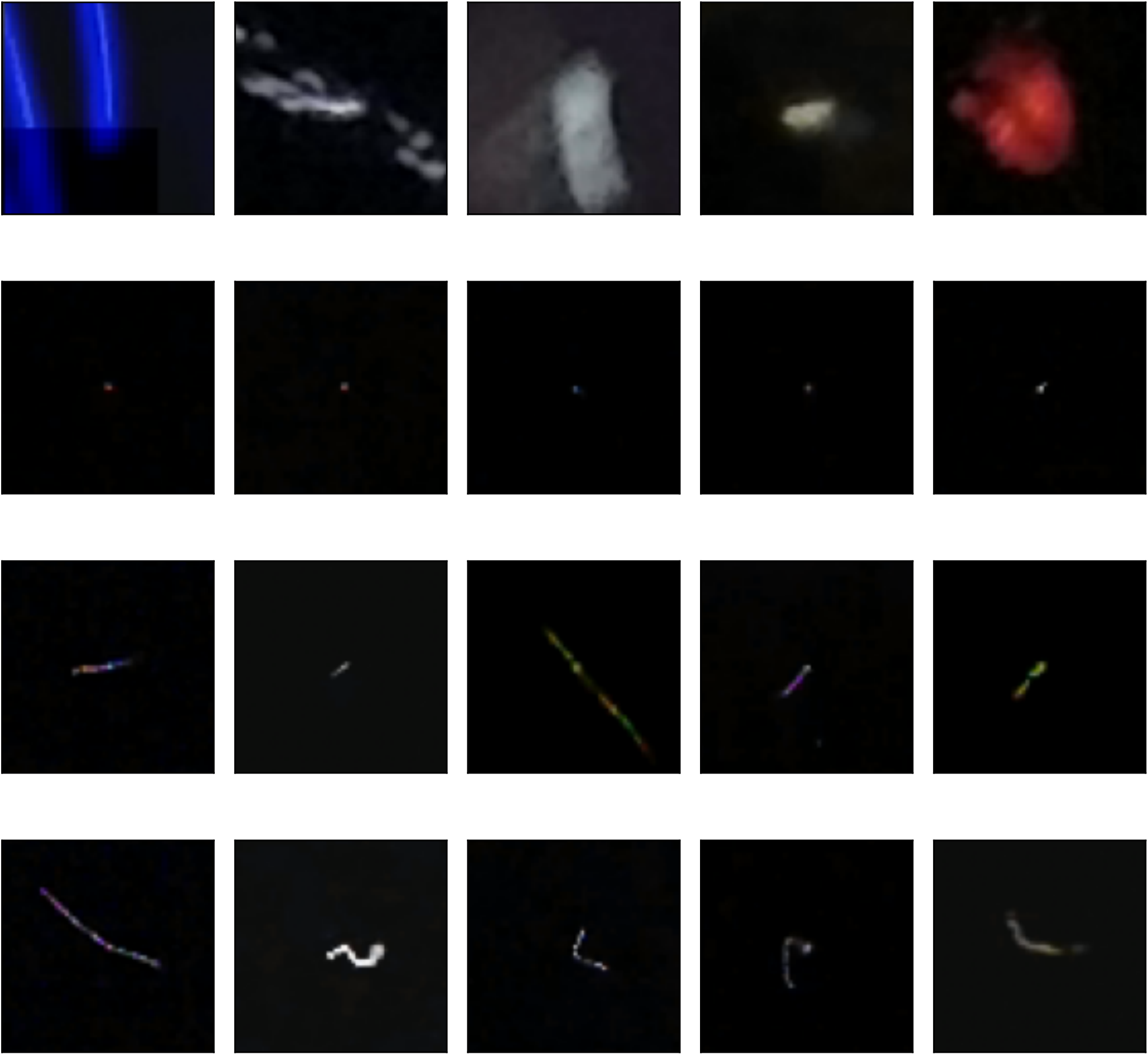}
    \caption{Examples of typical images labeled as artefacts (1st row), spots (2nd row), tracks (3rd row) and worms (4th row).}
    \label{fig:hits}
\end{figure}
One sees from Fig.~\ref{fig:hits} that there are indeed two characteristic features that distinguish artefacts and signals (spots, tracks and worms combined). These are the integrated brightness and the number of lit (active) pixels. Thus we define the decision boundary for the base trigger in terms of these two variables that we denote $\lambda$ and $\nu$, respectively. $\lambda$ and $\nu$ are computed relative to the adaptively computed threshold defined by Eq.~\ref{threshold}. 
\begin{equation}
\begin{array}{c}
threshold=\left\{
        \begin{array}{l}
        t_i\,\,\textrm{for}\,\, t_i<100\\
        100\,\,\textrm{for}\,\,t_i\ge 100,
        \end{array}
        \right.
\end{array}
\label{threshold}
\end{equation}
where $t_i$ is defined by
\begin{equation}
t_i=\overline{b}_i+5\sigma_i,
\label{stdev}
\end{equation}
with $\overline{b}_i$ and $\sigma_i$ denoting the average and standard deviation. 
For the artefact and signal classes we define the limiting values of $\lambda$ and $\nu$, namely the minimum integrated luminosity $\lambda_{min}^{art}$ and minimum number of active pixels for images labeled as artefacts $\nu_{min}^{art}$ and maximal integrated luminosity $\lambda_{max}^{sig}$ and maximal number of active pixels $\nu_{max}^{sig}$ for images labeled as signals. The  parameters determining the decision boundary are defined as the average of respective quantities $\lambda_b=(\lambda_{min}^{art}+\lambda_{max}^{sig})/2$ and
    $\nu_b=(\nu_{min}^{art}+\nu_{max}^{sig})/2$.
With parameters defining the decision boundary we define the decision boundary itself as the quarter ellipse
    \begin{equation}
        \left( \frac{\nu}{\nu_b} \right)^2+
        \left( \frac{\lambda}{\lambda_b} \right)^2=1,\quad \text{with}\quad \nu, \lambda>0.
        \label{eq:decision.ellipse}
    \end{equation}
Thus all hits whose integrated luminosity and number of active pixels fall inside the quarter ellipse are classified as signals and those outside of it, as artefacts. The distribution of the signal and artefact labelled examples in the $(\nu,\lambda)$ plane is shown in Fig.~\ref{fig:basetrigger}.
From Fig.~\ref{fig:basetrigger} one immediately sees that the decision boundary of the form given by Eq.~\ref{eq:decision.ellipse} indeed separates the the artefact and signal examples except for the small admixture of artefacts with small luminosities and number of active pixels. The accuracy obtained for the baseline trigger on the evaluation data set is surprisingly good and reaches almost 99\% for signals and 95\% for artefacts (see~\cite{CNNTrigger} for details). 
\section{Artefact rejection with the CNN based trigger}
Having defined the baseline trigger we can refine the artefact rejection process by employing a more elaborate classifier. To this end we use the Convolutional Neural Network trained on both the preprocessed and unpreprocessed  dataset. The preprocessing  phase is performed in two versions. In the first version the raw images are just subject to adaptive thresholding as per Eq.~\ref{threshold}. In the second version, apart from thresholding the Daubechies wavelet transform is applied~\cite{doi:10.1063/1.5139378}. 
After the preprocessing the images are fed to the CNN for training and evaluation (see~\cite{CNNTrigger} for details on the network architecture and evaluation).
The objective of the wavelet transform is to enhance the trigger capability to recognize the borders of the traces left by particles. The Daubechies wavelet transforms the original image into 4 images of two times smaller resolution than the original image according to the formula
\begin{equation}
\begin{gathered}
    \mathbf{f}\to
    \left(
    \begin{array}{ccc}
       \mathbf{a} & |  & \mathbf{v} \\
       -& & -\\
       \mathbf{h} & | &  \mathbf{d}
    \end{array}
    \right),
    \label{eq:wavelettransf}
\end{gathered}
\end{equation}
where $\mathbf{f}$ is the original image function, subimage $\mathbf{a}$ denotes the average signal while $\mathbf{h}$, $\mathbf{v}$ and $\mathbf{d}$ denote the horizontal, vertical and diagonal fluctuations, respectively \cite{Walker_1999}. The fluctuation components are usually decisive for detecting the objects' borders (trace border in this case). However, in this study we found that the impact of using either single or combined Daubechies wavelets does not impact the trigger performance.
\begin{figure}[h]
\centering
\includegraphics[scale=0.33]{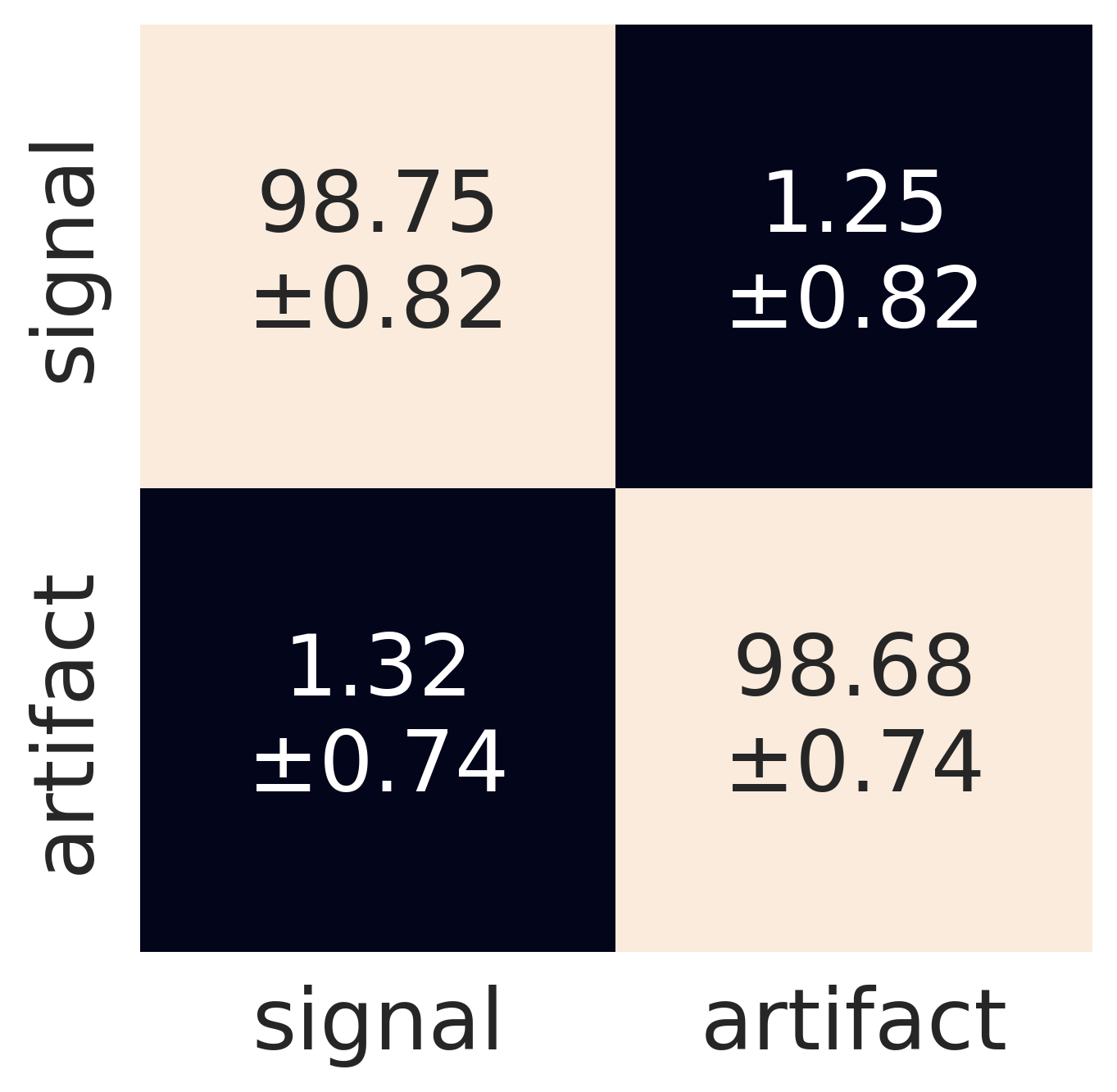} 
\includegraphics[scale=0.079]{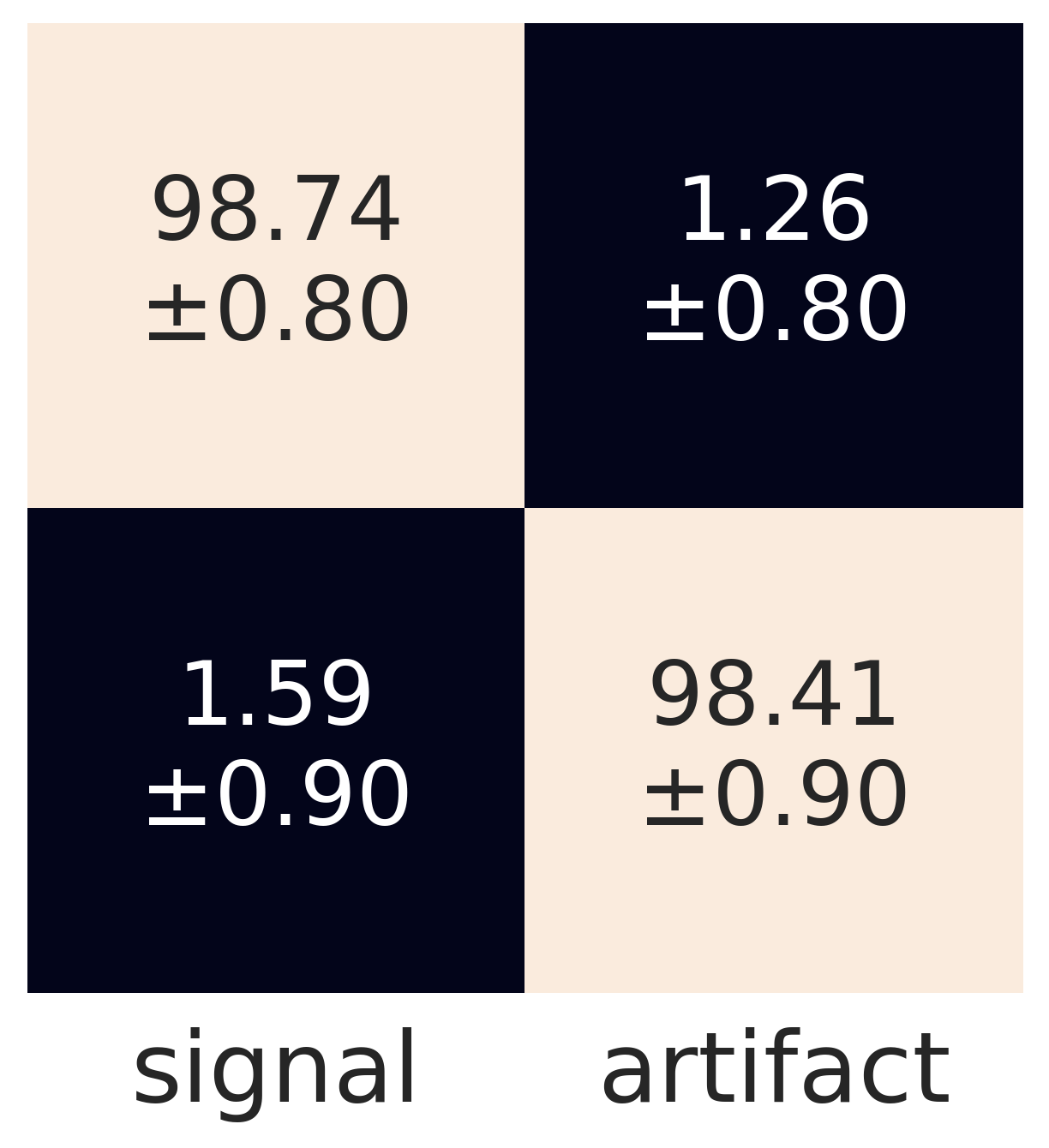} 
\includegraphics[scale=0.33]{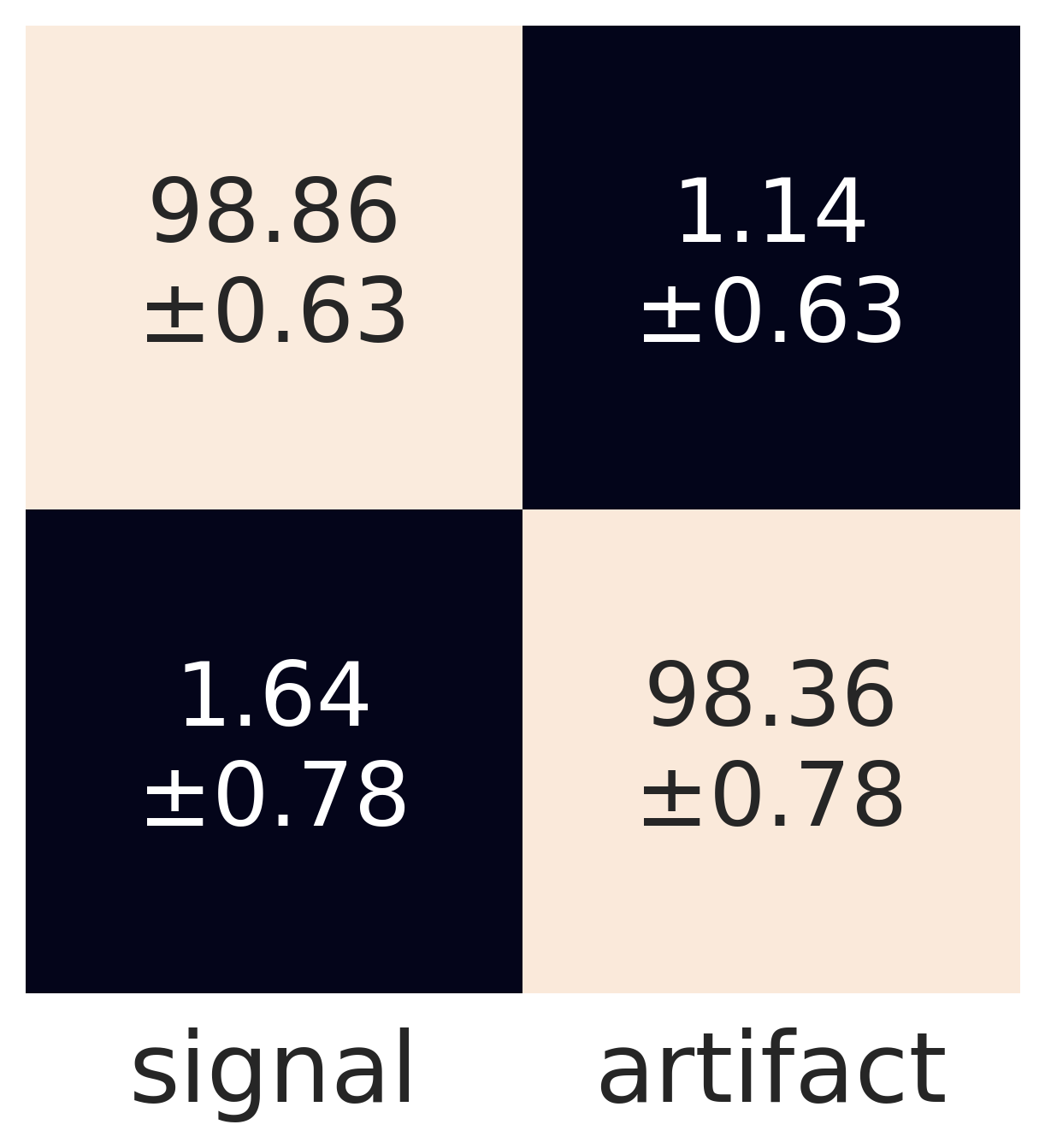} 
\includegraphics[scale=0.33]{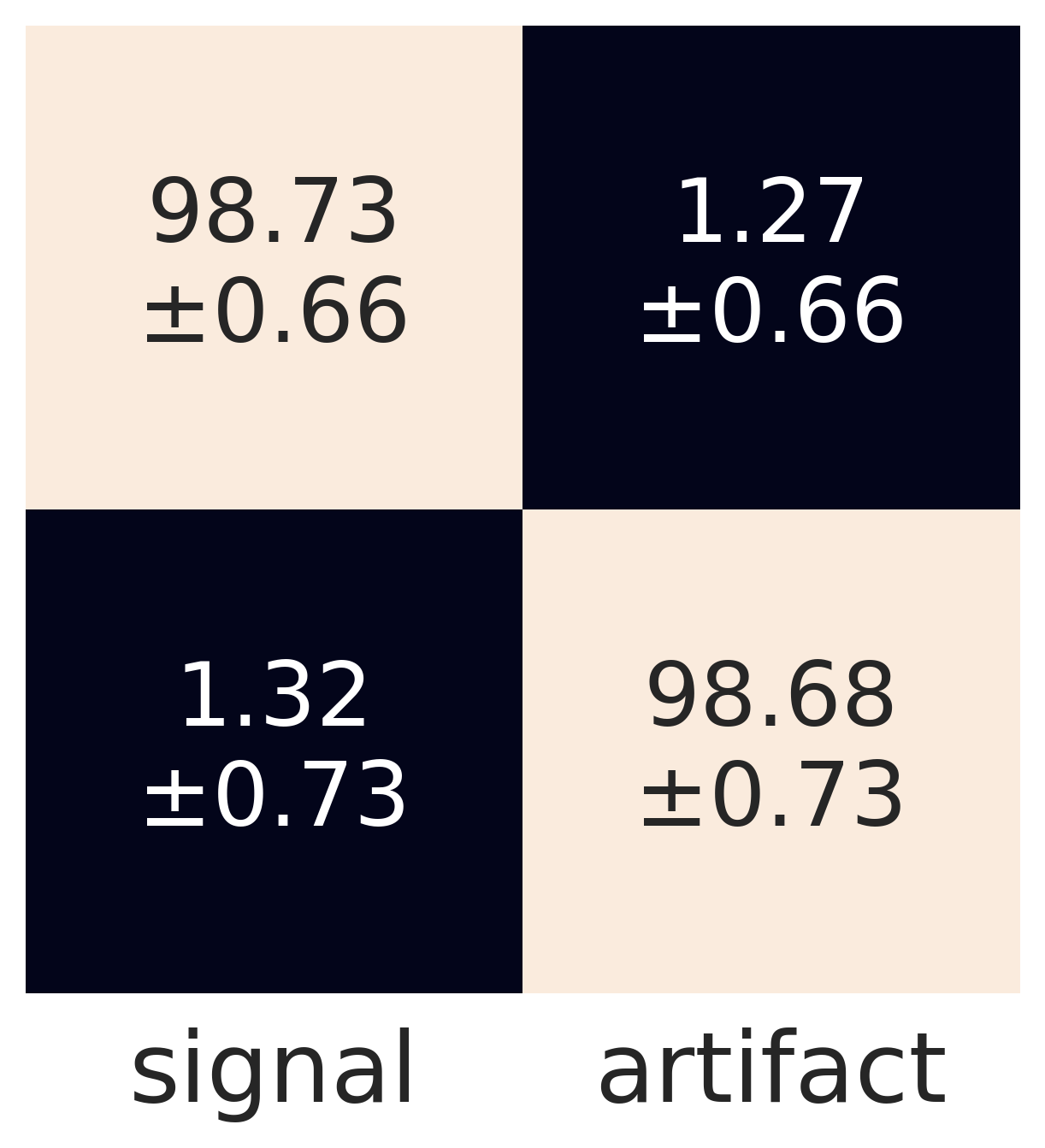} 
\caption{Confusion matrices for three configurations of input tensors (raw image, D2, D2:D4, D2:D20). The horizontal and vertical dimensions refer to predicted and annotated labels, respectively.}
\label{fig:cm-CNNtrigger}
\end{figure}
The overall trigger performance is illustrated by confusion matrices in Fig.~\ref{fig:cm-CNNtrigger}. It confirms that the signal-artefact separation accuracy is identical within one standard deviation, regardless of either applying a wavelet transform or not. This, quite surprising result which we interpret as a consequence of the fact that (as shown in Fig.~\ref{fig:hits}) both the signal and artefact classes are very diverse morphologically and combining examples with very diverse features, even though determined very accurately, leads to strong feature averaging. 
\section{Signal classification}
Having established a method of separating the images depicting signals and artefacts we went on to subdivide the signal class into three morphologically distinct classes of spots, tracks and worms. Together with artefacts this makes the problem of the four class categorical classification. In the literature these classes have been associated with particular particles species~\cite{Groom2002}, however, here we focus just on their morphological differences and leave particle association for further study.

Unlike the trigger where we had used the binary CNN classifier, in this exploratory study of the four class classification we employed a series of feature based classical statistical classifiers, like naive Bayesian, decision trees, kNNs, support vector machines, logistic regression, multilayer perceptrons as well as their ensemble versions. We used Zernike moments as feature carriers for all classifiers \cite{6460899,LAJEVARDI20101771,10.1007/978-3-319-60618-7_39}. Since the images to be classified were relatively simple compared to eg. human faces, applying the low order moments seemed to be justified. Therefore we used the $8^{th}$ order moments with 25 components. As the Zernike moments are known to be noises sensitive before computing them we had used the same noise cut-off procedure as in binary classification, defined by Eq.~\ref{threshold}. Then the components of Zernike moments were used train and evaluate the classifiers. We used two complementary methods of classifier evaluation - the 5-fold cross-validation and 30-fold evaluation on the randomly drawn (with stratification and replacement) 1/5 of the total data set. For a detailed discussion of applied classifiers and optimal values of respective hyperparameters see  
\cite{ZerClassifiers}. Here we summarize only the main results obtained with best performing classifiers that are shown in Table~\ref{tab:boosting_classifiers}.
\begin{table}[h]
\caption{Accuracy of the best performing ensemble classifiers. BAG+SVC - bagged support vector classifiers. OVO+MLP - multilier perceptrons with one vs. one decision function. OVO+SVC - support vector classifiers with one vs. one decision function. OVR+MLP multilier perceptrons with one vs. rest decision funtion.}
\label{tab:boosting_classifiers}
\centering
\begin{tabular}{ c c c c c }
\hline
\textbf{Classifier} & \textbf{CV} & \textbf{Test} & \textbf{Mean30} & \textbf{Std30} \\
\hline
BAG+SVC &	0.9078 &	0.8868 &	0.8805 &    0.0135  \\
OVO+MLP &	0.9101 &	0.8973 &	0.8880 &    0.0145  \\
OVO+SVC &	0.9171 &	0.8889 &	0.8850 &    0.0148  \\
OVR+MLP &	0.9138 &	0.8952 &	0.8853 &    0.0139  \\
\hline
\end{tabular}
\end{table}
The CV column in Table~\ref{tab:boosting_classifiers} contains the average classifier cross-validation accuracy in training phase where training involved 80\% of the original data set while test phase (Test column) involved the remaining 20\% of data. In order to guarantee the consistency across various classifiers they were trained and evaluated on the same data sets. Mean values Mean30 and standard deviations Std30 were computed by 30-fold drawing of the testing set that contained 20\% of examples in the whole data set.

\begin{figure}[h]
    \centering
    \vspace{-1cm}
    \includegraphics[scale=.5]{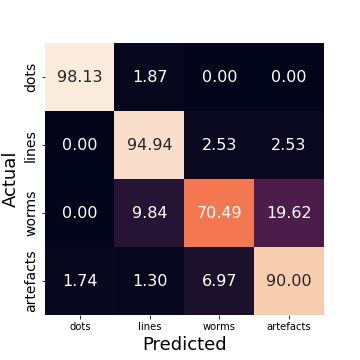}
    \caption{Four class confusion matrix for the best performing $\nu$-SVC classifier.}
    \vspace{-.2cm}
    \label{fig:four-class}
\end{figure}

To illustrate the overall performance across various trace classes we show in Fig.\ref{fig:four-class} the confusion matrix obtained for the best performing single (non-ensemble) $\nu$-SVC classifier. No surprise that the classifier is most confused to in recognizing worms vs. artefacts. It is also worth noting that the artefact recognition in the four-class scheme is worse than in the binary signal vs. artefact classification. This can be understood by inspecting Fig.~\ref{fig:basetrigger} where we see that some signals (most likely worms) are sitting within a decision boundary so are treated as "false artefacts". To put them back to their true signal class the classifier needs to penetrate the artefact zone in the ($\nu$,$\lambda$) plane but only at the cost of putting some artefacts in the false signal classes.
\section{Summary and outlook}
We have discussed the ANN based methods of the artefact rejection and hit classification applied to the data set containing candidate hits collected by the CREDO experiment. We have shown that the 99\% accuracy of the CNN trigger allows for entirely automatic handling of the surge of artefacts resulting from a wide use of gamification in this citizen science initiative. Further analysis performed with a wide array of statistical and ANN based classifiers enables the selection of the hits with particular morphologies. The overall accuracy of the four class signal classification on the test data set is about 88\% with the best recognition rate, reaching 98\% for spots and the worst - of about 70\% for worms. The classifiers are most likely to confuse worms and artefacts.

\acknowledgments{This research has been supported in part by PLGrid Infrastructure. We warmly thank the staff at ACC Cyfronet AGH-UST, for their always helpful supercomputing support.}


\bibliographystyle{JHEP}
\bibliography{CREDO-ML}

Full authors list (ONLY FOR COLLABORATIONS)
\clearpage
\section*{Full Authors List: \Coll\ Collaboration}




\scriptsize
\noindent
Lukasz Bibrzycki$^1$,
David Alvarez-Castillo$2$,
Olaf Bar$^1$,
Dariusz Gora$^2$,
Piotr Homola$^2$,
Péter Kovács$^3$,
Michał Niedźwiecki$^4$,
Marcin Piekarczyk$^1$,
Krzysztof Rzecki$^5$,
Jaroslaw Stasielak$^2$,
Sławomir Stuglik$^2$,
Oleksandr Sushchov$^2$ and
Arman Tursunov$^6$


\noindent
$^1$Pedagogical University of Krakow, Podchorążych 2, 30-084, Kraków, Poland\\
$^2$Institute of Nuclear Physics Polish Academy of Sciences, Walerego Eljasza Radzikowskiego 152, 31-342, Kraków, Poland\\
$^3$Wigner Research Centre for Physics, Konkoly-Thege Miklós út 29-33., H-1121, Budapest, Hungary\\
$^4$Faculty of Computer Science and Telecommunications, Cracow University of Technology, Warszawska 24, 31-155, Kraków, Poland\\
$^5$AGH University of Science and Technology, 30 Mickiewicza Ave., 30-059, Kraków, Polska\\
$^6$Institute of Physics, Silesian University in Opava, Bezručovo nám 13 , CZ-74601 , Opava, Czech Republic

\end{document}